\documentclass[aps,prl,showpacs,twocolumn,amsmath,10pt,superscriptaddress,floatfix,nofootinbib,a4paper]{revtex4-1}
\usepackage{epsfig,amssymb,amsfonts,bm,color}

\newcommand{\be}{\begin{equation}}
\newcommand{\ee}{\end{equation}}
\newcommand{\bea}{\begin{eqnarray}}
\newcommand{\eea}{\end{eqnarray}}
\newcommand{\beas}{\begin{eqnarray*}}
\newcommand{\eeas}{\end{eqnarray*}}

\newcommand{\jp}{J/\psi p}
\newcommand{\ld}{\Lambda_c^+ \bar D^0}
\newcommand{\ldstar}{\Lambda_c^+\bar D^{*0}}
\newcommand{\lag}{\mathcal{L}}
\newcommand{\amp}{\mathcal{A}}
\newcommand{\lrp}{{\stackrel{\leftrightarrow}{\partial}}}

\newcommand{\bonn}{\affiliation{Helmholtz-Institut f\"ur Strahlen- und Kernphysik and Bethe Center for Theoretical Physics,\\ Universit\"at Bonn, D-53115 Bonn, Germany}}

\newcommand{\fzj}{\affiliation{Institute for Advanced Simulation, Institut f\"ur Kernphysik and J\"ulich Center for Hadron Physics,\\ Forschungszentrum J\"ulich, D-52425 J\"ulich, Germany}}

\newcommand{\itp}{\affiliation{CAS Key Laboratory of Theoretical Physics, Institute of Theoretical Physics, Chinese Academy of Sciences,\\ Zhong Guan Cun East Street 55, Beijing 100190, China}}

\newcommand{\ucas}{\affiliation{School of Physical Sciences, University of Chinese Academy of Sciences, Beijing 100049, China}}

\newcommand{\itep}{\affiliation{{Institute for Theoretical and Experimental Physics NRC ``Kurchatov Institute'', Moscow 117218, Russia  }}}

\newcommand{\lebedev}{\affiliation{P.N. Lebedev Physical Institute of the Russian Academy of Sciences, 119991, Leninskiy Prospect 53, Moscow, Russia}}

\newcommand{\nrnu}{\affiliation{National Research Nuclear University MEPhI, 115409, Kashirskoe highway 31, Moscow, Russia}}

\newcommand{\mipt}{\affiliation{Moscow Institute of Physics and Technology, 141700, Institutsky lane 9, Dolgoprudny, Moscow Region, Russia}}

\newcommand{\tb}{\affiliation{Tbilisi State University, 0186 Tbilisi, Georgia}}

\newcommand{\GW}{\affiliation{Institute for Nuclear Studies, Department of Physics, The George Washington University, Washington, D.C. 20052, USA}}

\graphicspath{{Figs.dir/}}

\begin{document}

\title{Deciphering the mechanism of near-threshold $J/\psi$ photoproduction}

\author{Meng-Lin Du}
\bonn

\author{Vadim Baru}
\bonn \itep \lebedev

\author{Feng-Kun Guo}
\itp \ucas

\author{Christoph~Hanhart}
\fzj

\author{Ulf-G. Mei{\ss}ner}
\bonn\fzj\tb

\author{Alexey Nefediev}

\lebedev
\mipt
\nrnu

\author{Igor Strakovsky}
\GW

\begin{abstract}
The photoproduction of the $J/\psi$ off the proton is believed to deepen our understanding of various
physics issues.  On the one hand, it is proposed to provide access to the origin of the proton mass, based
on the QCD multipole expansion.
On the other hand, it can be employed in a study of pentaquark states. The process is usually assumed
to proceed through vector-meson dominance, that is the photon couples to a $J/\psi$ which rescatters
with the proton to give the $J/\psi p$ final state. In this Letter, we provide a compelling hint
for and propose measurements necessary to confirm a novel production mechanism via the
$\Lambda_c \bar D^{(*)}$ intermediate states. In particular, there must be cusp structures at the $\Lambda_c \bar D^{(*)}$ thresholds in the energy dependence of the $J/\psi$ photoproduction cross section.
The same mechanism also implies the $J/\psi$-nucleon scattering lengths of order 1~mfm. Given this,
one expects only a minor contribution of charm quarks to the nucleon mass.
\end{abstract}

\maketitle

\section{Introduction}

\begin{figure}[tb]
\centerline{\includegraphics[width=0.6\columnwidth]{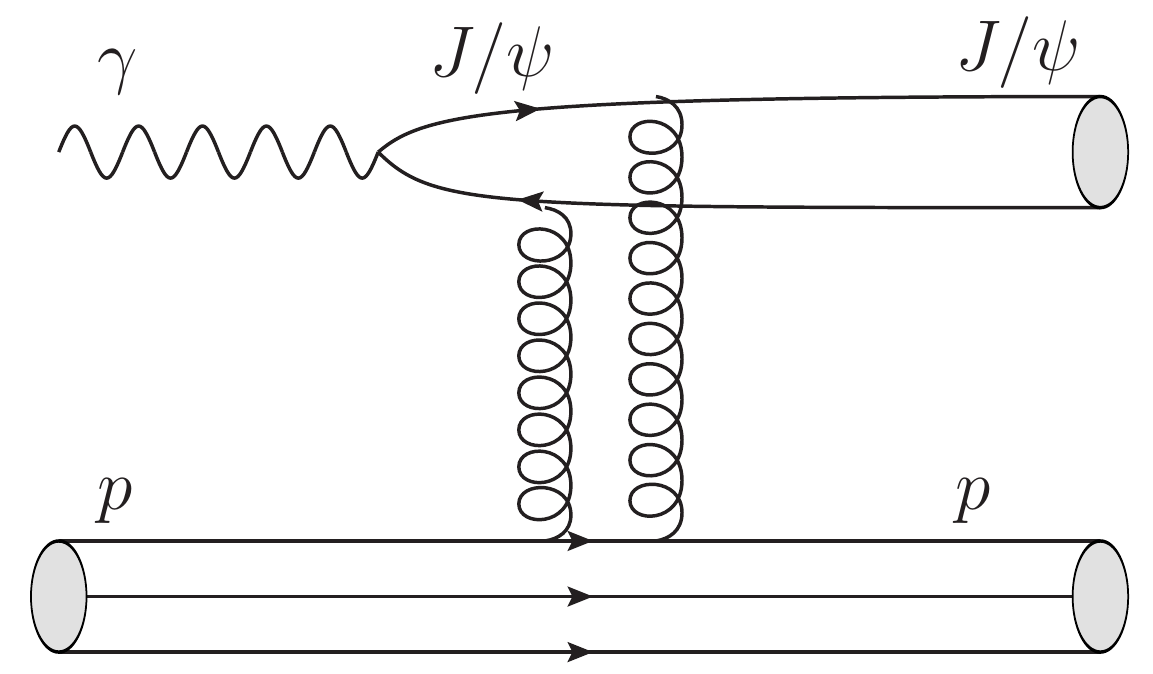}}
\caption{Vector-meson dominance model mechanism for the near-threshold $J/\psi$ photoproduction.}
\label{fig:mechanismvdm}s
\end{figure}

Understanding how strong interactions work in the nonperturbative regime of quantum chromodynamics (QCD)
remains one of the most challenging tasks of the Standard Model. One fundamental problem tied to the
nonperturbative nature of QCD is how the visible matter of the universe gets most of its mass that can
be translated to how the proton and neutron---the fundamental ingredients of all kinds of nuclei
in the universe---acquire their masses. 
It was suggested that the near-threshold production of heavy quarkonium is sensitive to the trace
anomaly contribution to the nucleon mass~\cite{Kharzeev:1995ij,Kharzeev:1998bz} which may be measured
at the Jefferson Laboratory and future electron-ion colliders~\cite{Gryniuk:2020mlh} (for recent
discussions see, for example, Refs.~\cite{Hatta:2018ina,Wang:2019mza}). This suggestion is based on
the vector-meson-dominance (VMD) model and the assumption that the nucleon interacts with a heavy
quarkonium through multiple-gluon exchange, as illustrated in Fig.~\ref{fig:mechanismvdm}.  
If this is indeed the dominant mechanism, the near-threshold $J/\psi$ photoproduction cross section
would provide the  link to the $J/\psi p$ elastic scattering amplitude at low energies which is
fundamentally important because  the $J/\psi p$ scattering length can be related to the nucleon matrix
element of two-gluon operators and thereby to the trace anomaly contribution to the nucleon mass.
Also, a possible existence of the quarkonium-nucleus bound states first proposed in
Ref.~\cite{Brodsky:1989jd}    crucially depends on the strength of $J/\psi N$ interaction at low
energies, characterised  by the $J/\psi p$ scattering length. 
A loophole with this mechanism is, however, that it relies on the QCD multipole expansion, see below.

Another fundamental issue of nonperturbative QCD is that it is still unclear how the hadron spectrum
is organized. In particular, exotic hadrons such as multiquark states are being sought experimentally
and studied theoretically using phenomenological models, effective field theories and lattice QCD. In the
last two decades, tens of states beyond the conventional quark model were found, many of them by different
experiments in different reactions. However, their structure still needs to be resolved. For recent reviews
on both theoretical and experimental aspects of exotics in the heavy quark sector see, for example,
Refs.~\cite{Hosaka:2016pey,Lebed:2016hpi,Esposito:2016noz,Guo:2017jvc,Olsen:2017bmm,Liu:2019zoy,Brambilla:2019esw,Guo:2019twa,Yang:2020atz}.
An intriguing recent discovery was a set of hidden-charm pentaquark candidates, observed in the $\Lambda_b$ decays
by the LHCb Collaboration~\cite{Aaij:2015tga,Aaij:2019vzc}  and called $P_c$ states, which triggered a flood of
theoretical investigations. However, a subsequent search of the $P_c$ states in the 
GlueX experiment using the photoproduction process $\gamma p\to J/\psi p$ did not reveal any
signal~\cite{Ali:2019lzf}. 
The analyses in Refs.~\cite{Cao:2019kst,Winney:2019edt} which conclude that the branching fraction of the
$P_c\to J/\psi p$ should be at most a few per cent are also based on the VMD model: The photon is
assumed to convert to a $J/\psi$ which rescatters then with the proton target to form $P_c$ states.
In fact, VMD is generally assumed in estimating the cross sections for the photoproduction of
hidden-charm and hidden-bottom pentaquark states~\cite{Kubarovsky:2015aaa,Karliner:2015voa,Huang:2016tcr,Meziani:2016lhg,Paryev:2018fyv,Blin:2016dlf,Paryev:2018fyv, Cao:2019kst,Wang:2019krd,Wu:2019adv,Winney:2019edt,Cao:2019gqo,Yang:2020eye,Paryev:2020jkp}.

The rich physical implications related to the photoproduction of the $J/\psi$ off the proton provide a
strong motivation to revise the assumptions underlying the VMD  approach, and identify its  possible caveats.  
Specifically, (i) the $J/\psi$ attached to the photon is highly off-shell while the $J/\psi p$ scattering
length is defined for the on-shell scattering amplitude; (ii) 
the $\Lambda_c^+\bar D^0$ threshold is only 116~MeV above the $J/\psi p$ threshold, rendering the contribution
from the $\Lambda_c \bar D$ channel potentially 
sizeable and thus making the relation between the photoproduction cross section and the trace anomaly
contribution to the nucleon mass even more obscure. 
In this Letter, we investigate the implications of the latter observation.
\begin{figure}[tb]
\centerline{\includegraphics[width=0.6\columnwidth]{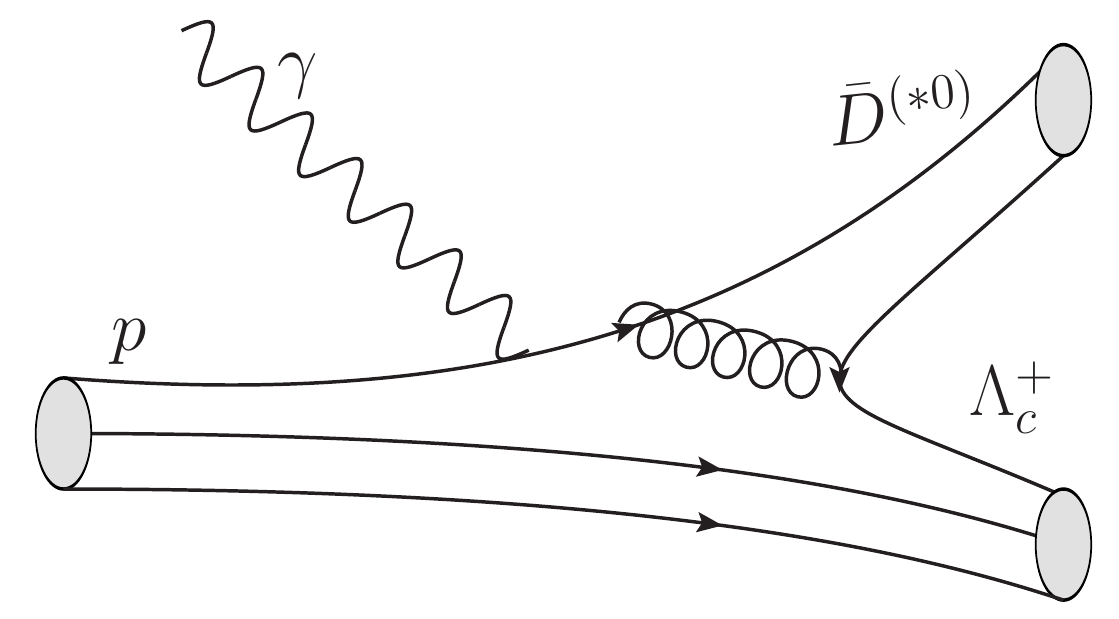}}
\caption{Mechanism for the near-threshold $J/\psi$ photoproduction through $\Lambda_c\bar D^{(*)}$ which then rescatter into $J/\psi p$.}
\label{fig:mechanismcc}
\end{figure} 

We propose a new coupled-channel (CC) mechanism for the near-threshold $J/\psi$ photoproduction which
is not directly related to the nucleon matrix element of the gluonic operator since the $J/\psi p$ final
state is produced through the nearby open-charm channels $\Lambda_c\bar D$ and $\Lambda_c\bar D^*$, see
Fig.~\ref{fig:mechanismcc}. In particular, we demonstrate that the data recently measured at GlueX
can be quantitatively understood using this mechanism with reasonable parameters.  
With this mechanism, the direct relation between the trace anomaly contribution to the nucleon
mass and the $J/\psi$ near-threshold photoproduction, that is present in the VMD model, is obscured.
We discuss the implications  of this mechanism,  and  suggest experimental observables which should allow
one to test the picture outlined here.

\section{Coupled-channel mechanism}
\begin{figure}[tb]
\centerline{\includegraphics[width=0.3\textwidth]{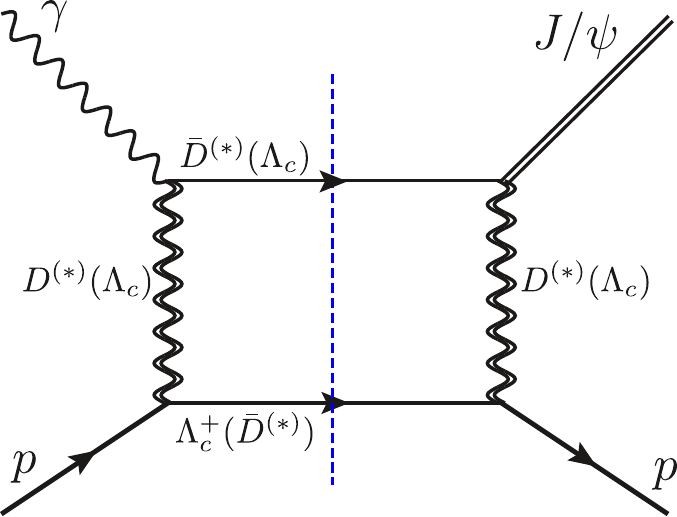}}
\caption{Feynman diagram for the proposed CC mechanism. The dashed blue line pinpoints the
  open-charm intermediate state.}
\label{fig:feyngraph}
\end{figure}

The cross section for the inclusive production of a charm and anti-charm quark pair, $\gamma p\to c\bar c X$
with $X$ denoting everything that is not detected, is about two orders of magnitude higher than that for
the exclusive production of the $J/\psi$, $\gamma p\to J/\psi p$ (for a compilation of the data and a
VMD model fit see Ref.~\cite{Gryniuk:2016mpk}). This might indicate that the cross sections for the pairs
of open-charm mesons and baryons are sizeable, which was also expected in Ref.~\cite{Boreskov:1992ur}.
Then, open-charm channels close to the $\jp$ threshold could potentially contribute significantly to
the $\jp$ production. While there are no data for the photoproduction of open-charm channels in the pertinent
energy region yet, it should be noted that the cross sections for the analogous 
reactions in the strangeness sector, $\gamma p\to K^+\Lambda/K^+\Sigma^0$~\cite{Bradford:2005pt,Bradford:2006ba,McCracken:2009ra,Dey:2010hh,Paterson:2016vmc},
are indeed much larger than that for the near-threshold $\phi$ meson production,
$\gamma p\to \phi p$~\cite{Seraydaryan:2013ija,Dey:2014tfa,Mizutani:2017wpg}. 

For the $\jp$ photoproduction off the proton, the closest open-charm channels are $\ld$ and $\ldstar$
with the thresholds just 116 and 258~MeV above the $\jp$ threshold, respectively. In this Letter, we
investigate the contribution of these channels to the $\jp$ photoproduction. Not only is the cross section
estimated by considering the exchange of $D^{(*)}$ and $\Lambda_c^+$, as shown in Fig.~\ref{fig:feyngraph},
but also are general features of the resulting rates identified and possible future experiments suggested to
test the proposed mechanism.

In the near-threshold region, it is sufficient to consider effective Lagrangians that have the smallest
number of derivatives, which are given as follows,
\begin{eqnarray}
\lag_{\Lambda_c D N} &=& - g_{D^*N\Lambda_c} \bar{\Lambda}_c \gamma_\mu N D^{*\mu} - i g_{DN\Lambda_c} \bar{\Lambda}_c \gamma_5 N D \nonumber\\  
&& - g_{D^*N\Lambda_c} \bar{N}\gamma_\mu \Lambda_c D^{*\mu\dag} - i g_{DN\Lambda_c} \bar{N}\gamma_5\Lambda_c D^\dag , \label{lag:LDN}\\
\lag_{\psi} &=& - g_{\psi DD^*} \psi_\mu \epsilon_{\mu\nu\alpha\beta}\big( \partial_\nu D_\alpha^*\partial_\beta D^\dag -\partial_\nu D\partial_\beta D_\alpha^{*\dag} \big), \nonumber\\
& &+ ig_{\psi D^*D^*}\psi^\mu\big( D^{*\nu}\partial_\nu D_\mu^{*\dag} -\partial_\nu D_\mu^* D^{*\nu\dag} \nonumber\\ 
&& -D^{*\nu} \lrp_\mu D_\nu^{*\dag}\big) -ig_{\psi DD} D^\dag \lrp_\mu D \psi^\mu \nonumber\\
&&
+ g_{\psi \Lambda_c \Lambda_c}\bar{\Lambda}_c\gamma_\mu \psi^\mu\Lambda_c, \label{lag:PsiDD}\\
\lag_\gamma  &=& - g_{\gamma DD^{*}} F_{\mu\nu}\epsilon^{\mu\nu\alpha\beta} (D^{*}_\alpha {\stackrel{\leftrightarrow}{\partial}}_\beta D^\dag - D{\stackrel{\leftrightarrow}{\partial}}_\beta D^{*\dag}_\alpha ) \nonumber\\ 
&&- i g_{\gamma D^*D^*} F^{\mu\nu} D^{*\dag}_\mu D^{*}_\nu -e \bar{\Lambda}_c\gamma_\mu A^\mu \Lambda,\label{lag:gammaDD}
\end{eqnarray}
where $D$ and $D^*$ refer to the fields for the neutral charmed mesons, $e$ is the elementary (positive) electric charge ($\alpha=e^2/(4\pi)\simeq 1/137$) and the
couplings $g_{\psi DD}=g_2 m_D\sqrt{m_\psi}$, $g_{\psi DD^*} = g_2\sqrt{m_\psi m_D/m_{D^*}}$, $g_{\psi D^*D^*} =
g_2m_{D^*}\sqrt{m_\psi}$ are related to the same coupling constant $g_2$ through heavy
quark spin symmetry~\cite{Colangelo:2003sa,Guo:2010ak}.  
\begin{table*}[tb]
\caption{Values of the couplings in the Lagrangians in Eqs.~\eqref{lag:LDN}-\eqref{lag:gammaDD} used in the calculation.}
\begin{ruledtabular}
\begin{tabular}{l|ll|lll|l}
Coupling & $g_{\gamma DD^*}$& $g_{\gamma D^*D^*}$ &   $g_{DN\Lambda_c}$ & $g_{D^*N\Lambda_c}$ & $g_{\psi\Lambda_c\Lambda_c}$ & $g_{\psi DD}$ \\
\hline
Value & 0.134~GeV$^{-1}$& 0.641 & $-4.3$ &  $-13.2$ & $-1.4$   &7.44 \\
\hline
Source & \multicolumn{2}{l|}{Experimental data \cite{Zyla:2020zbs}} &  \multicolumn{3}{l|}{\hspace*{2.cm} SU(4) \cite{Liu:2001yx,Oh:2007ej}} & \hspace*{-5mm} {VMD \cite{Liu:2001yx,Oh:2007ej}}
\end{tabular}
\end{ruledtabular}
\label{tab:params}
\end{table*}

Since electric charge conservation law allows only for a contribution of the neutral $D^{(*)}$ mesons
(see Fig.~\ref{fig:feyngraph}), the Lagrangian $\gamma D^{(*)} D^{*}$ in Eq.~(\ref{lag:gammaDD}) contains
only magnetic interactions. The corresponding couplings can be fixed directly from the data on the
experimentally measured total width of the $D^{*+}$ meson (the unknown total width of the $D^{*0}$ meson is
evaluated using isospin symmetry) and the branching fraction of the decay $D^{*0}\to D^0\gamma$~\cite{Zyla:2020zbs}.
For the other couplings we employ predictions of phenomenological approaches, the corresponding values are
collected in Table~\ref{tab:params}. We notice that using the couplings $g_{DN\Lambda_c}=-10.7$ and
$g_{D^*N\Lambda_c}=-5.8$ obtained from the light-cone sum rules \cite{Khodjamirian:2011jp,Khodjamirian:2011sp}
give similar results which we therefore do not quote here.

\section{Comparison with the data}

The amplitude for the box diagram from Fig.~\ref{fig:feyngraph} is evaluated using a dispersion relation as 
\be
    \frac{1}{\pi}\int_{\text{th}}^{s_\text{cut}} \frac{\amp_{\gamma p \to \Lambda_c^+\bar{D}^{(*)0} }(s^\prime)\rho(s^\prime)
    \amp_{J/\psi p \to \Lambda_c^+\bar{D}^{(*)0}}(s^\prime) }{s^\prime - s}ds^\prime,
    \label{eq:estimationcut}
\ee
with $\text{th} = (m_{\Lambda_c}+m_{\bar{D}^{(*)}})^2$, where both amplitudes ${\cal A}$ involved are worked
out using the Lagrangians (\ref{lag:LDN})-(\ref{lag:gammaDD}), and $\rho=q_{\rm cm}/(8\pi\sqrt{s})$ is
the two-body phase space with $\sqrt{s}$ and $q_{\rm cm}$ the energy and the magnitude of the three-momentum
in the center-of-mass frame, respectively. The dispersive integral in Eq.~(\ref{eq:estimationcut}) is cut off at 
\be
    \sqrt{s_{\rm cut}}=\sqrt{q_{\rm max}^2+m_{\Lambda_c}^2}+\sqrt{q_{\rm max}^2+m_{D}^2},
\ee 
with a natural value for $q_{\rm max}$ being about 1~GeV.
Only the contribution of the $S$ wave is retained for the open-charm system $\bar{D}^{(*)}\Lambda_c$ near
threshold while both $S$ and $D$ waves are considered for the $J/\psi p$ and $\gamma p$ systems.

To take into account that the exchanged particles (doubly-wavy lines in Fig.~\ref{fig:feyngraph}) are
off-shell with a potentially large virtuality, we augment them with a single-pole form factor~\cite{Colangelo:2003sa,Gortchakov:1995im,Cheng:2004ru},
\be
    F(t)=\frac{\Lambda^2-m_{\text{ex}}^2}{\Lambda^2-t},
    \label{eq:ff}
\ee
which is consistent with the QCD counting rules~\cite{Colangelo:2003sa,Gortchakov:1995im}. A natural
value for the cutoff $\Lambda$ is the mass of the lowest neglected exchange particle, so that we set \cite{Cheng:2004ru}
\be
    \Lambda=m_\text{ex}+\eta\Lambda_\text{QCD},
    \label{eq:lambda}
\ee
where $\Lambda_{\rm QCD}\simeq 250$~MeV and the parameter $\eta$ which depends on both exchanged and external
particles \cite{Cheng:2004ru} is expected to be of order unity. For simplicity, if not stated otherwise,
we set $\eta=1$ and $\Lambda_{\rm QCD}=250$~MeV for all exchanged particles.
\begin{figure}
\begin{center}
\includegraphics[width=0.49\textwidth]{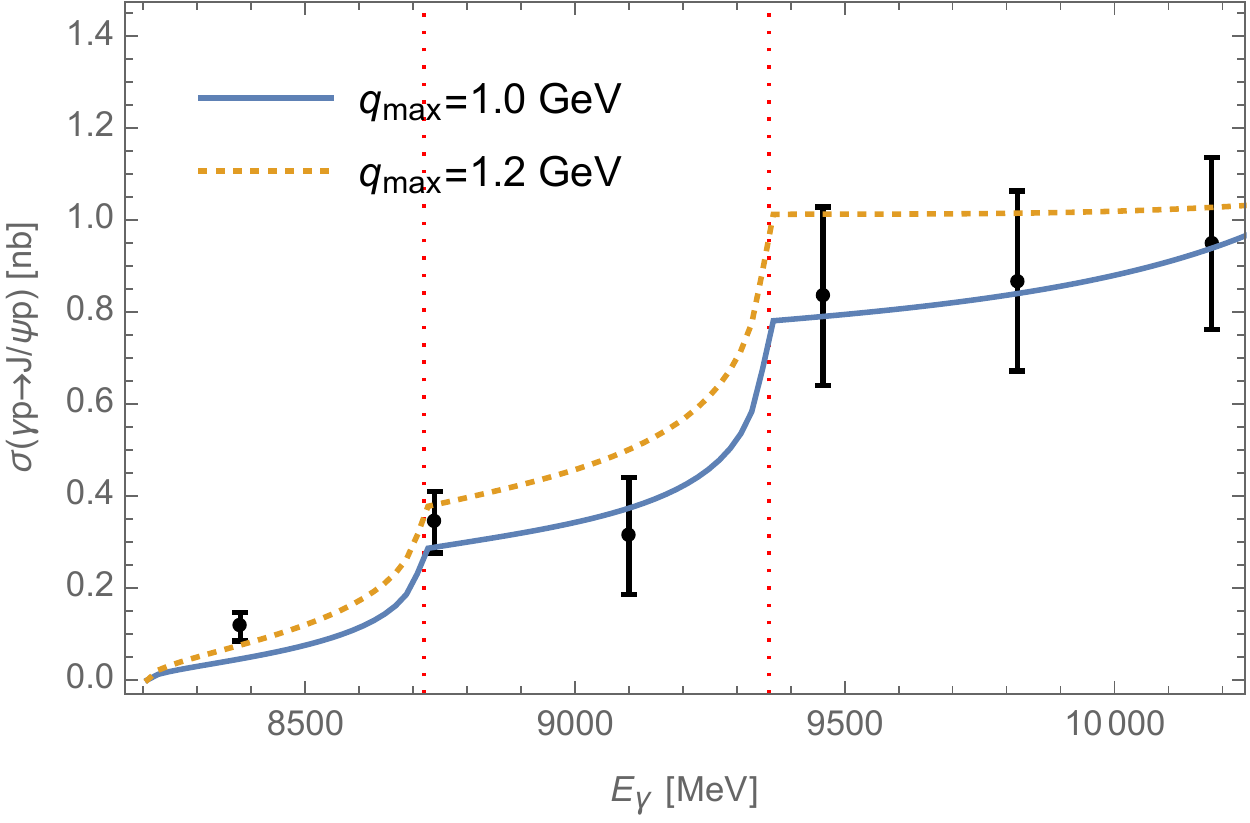}
\end{center}
\caption{Comparison of the $J/\psi$ photoproduction through the open-charm loops as shown in
  Figs.~\ref{fig:mechanismcc}  and \ref{fig:feyngraph} with the GlueX data~\cite{Ali:2019lzf}. $E_\gamma$ is the photon energy in the rest frame of the initial proton.
  Since we consider only the $\Lambda_c\bar D^{(*)}$ channels, the comparison with the data is
  only shown up to $E_\gamma=10.2$~GeV though  a qualitative agreement up to the highest GlueX data point 11.6~GeV
  is also achieved. The vertical dotted lines indicate the $\Lambda_c\bar D^{(*)}$ thresholds.
}\label{fig:section}
\end{figure}

The cross section as a function of the photon energy calculated using the amplitude (\ref{eq:estimationcut})
with the parameters from Table~\ref{tab:params} is shown in Fig.~\ref{fig:section} in comparison with the
data. No parameter is fitted or fine-tuned. Although the approach used suffers from several uncertainties
(badly determined couplings and form factors and only a limited set of diagrams considered to be mentioned
in the first place), we notice that not only does the cross section we obtain appear to have the right
order of magnitude but it also demonstrates a shape compatible with the data. We therefore dare to conclude
that the open-charm loop mechanism advocated here does 
indeed have the opportunity to make an important, possibly dominating, contribution to the $J/\psi$
photoproduction off the nucleon.  

\section{Predictions and possible tests}

We collect several immediate predictions of the mechanism discussed in this Letter, and enumerate
further experimental tests which should allow either to consolidate or falsify the picture outlined here.

\subsection{Threshold cusps}

The hypothesis that the suggested production mechanism through charmed intermediate states indeed
dominates the $J/\psi$ production leads to a unique prediction that can be verified in near-future experiments:
There must be sizeable cusps at the $\Lambda_c\bar D$ and $\Lambda_c\bar D^*$ thresholds. 
This is a universal phenomenon for $S$-wave thresholds, and the cusp shape is a measure of the strength
of the transition leading to the cusp (for a recent review of cusps in hadronic reactions, see Ref.~\cite{Guo:2019twa}).
Consequently, in the present data shown in Fig.~\ref{fig:section}, one is tempted to interpret a relatively
low cross section at $E_\gamma=9.1$~GeV as an indication of a nontrivial energy dependence of the cross
section near an open-charm meson-baryon threshold. The presence of such cusps as a clear indication of
the importance of the charm loops is a central finding of this Letter. 

\subsection{Production of open-charm final states}

\begin{figure}
\begin{center}
\includegraphics[width=0.49\textwidth]{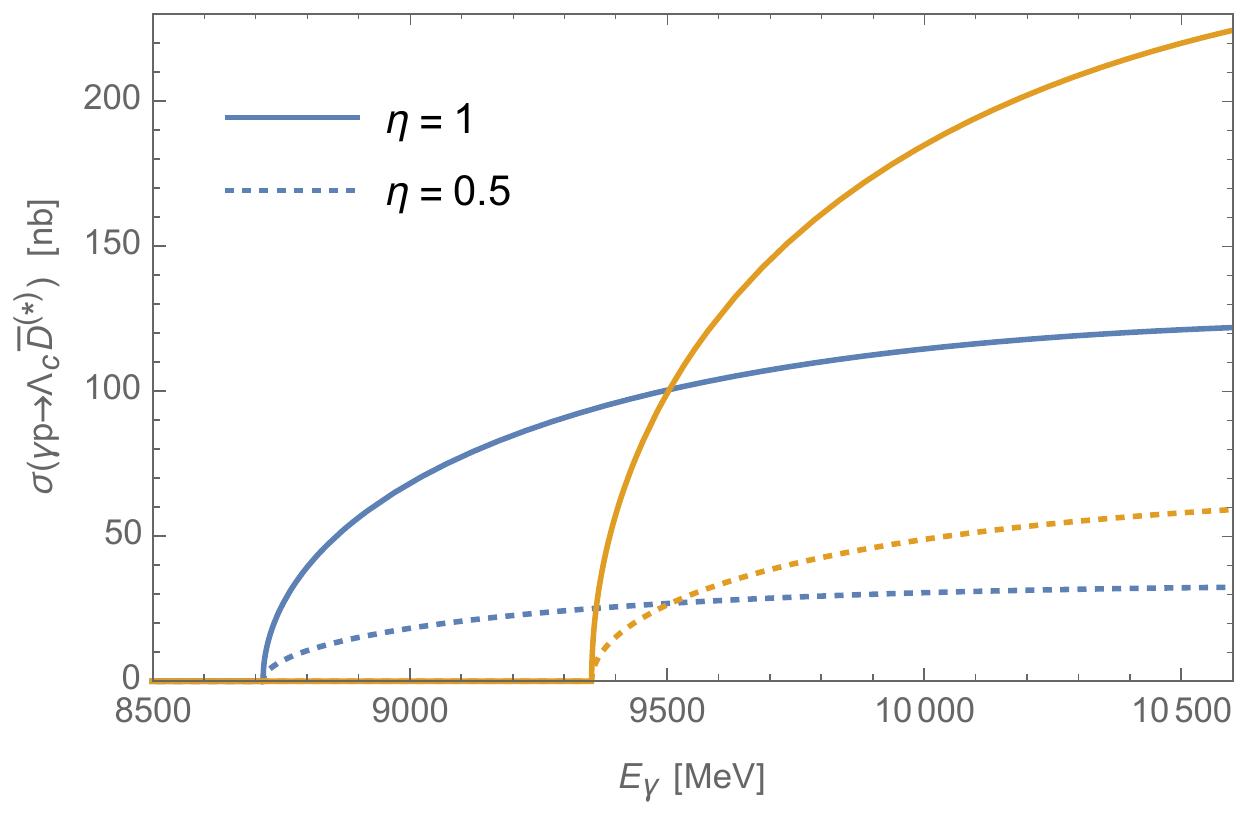}
\end{center}
\caption{Estimates of the cross sections for the $\gamma p\to \Lambda_c\bar{D}$ (blue curves) and
  $\gamma p\to \Lambda_c\bar{D}^{*}$ (orange curves) reactions. 
\label{fig:LambD}}
\end{figure}

Within the model advocated in this work we are in a position to provide an order-of-magnitude estimate
(neglecting the fine cusp structure that should also be present at the $\Lambda_c\bar D^*$ threshold)
for the not yet measured reactions $\gamma p\to \Lambda_c\bar D^{(*)}$, see Fig.~\ref{fig:LambD}. 
As an illustration of the sensitivity to the form factor, we show the results for $\eta=0.5$ and $1$.

The cross sections of the $\gamma p\to \Lambda_c \bar D^{(*)}$ reactions were calculated in Ref.~\cite{Wu:2019adv}
considering exchanges of $s$-channel  hidden-charm pentaquarks and $t$-channel $D^*$ mesons using the VMD
model. The corresponding predictions appear an order of magnitude smaller than those presented in
Fig.~\ref{fig:LambD}, which provides an additional support for the importance of the open-charm mechanism
suggested in this work.

\subsection{$J/\psi$-nucleon scattering lengths}

The suggested approach can be employed to evaluate the $J/\psi$-nucleon scattering lengths, replacing the photon
by a $J/\psi$ in Fig.~\ref{fig:feyngraph}. The results then appear to have the order of several units
of mfm. In particular, varying the parameter $\eta$, which affects this observable most strongly,
between $\eta=0.5$ and $\eta=2$, we find
\be
    \left|a^{J=1/2}\right|=0.2\dots 3.1~\text{mfm},~~\left|a^{J=3/2}\right|=0.2\dots 3.0~\text{mfm},
\ee
where $J$ corresponds to the total angular momentum of the $J/\psi$-nucleon system.  These numbers
are comparable with the recent estimation of the $J/\psi p$ scattering length  from the GlueX data using
the VMD model~\cite{Strakovsky:2019bev,Pentchev:2020kao} but much smaller than the results of the  two-gluon
exchange calculation using the multipole expansion~\cite{Luke:1992tm,Brodsky:1997gh},  for
a summary of the results  from  other calculations we refer to Ref.~\cite{Pentchev:2020kao}. 
The interactions between a nucleon and a quarkonium have also been studied on lattice, e.g., in
Refs.~\cite{Skerbis:2018lew,Sugiura:2017vks,Beane:2014sda,Liu:2008rza}. In the most recent lattice QCD
calculation of $J/\psi N$ scattering  of Ref.~\cite{Skerbis:2018lew}, the  lattice spectra in the
one-channel approximation  were found to be consistent with an almost non-interacting $J/\psi N$ system.
As stressed in Ref.~\cite{Skerbis:2018lew}, the lattice results suggest that the existence of the $P_c$
resonances  within a one-channel $J/\psi N$ scattering  is not favored in QCD and that the strong
coupling between the $NJ/\psi$ with other two-hadron channels might be responsible for the existence of
the $P_c$ resonances. This conclusion is in line with the current analysis.

\section{Summary}

In this Letter, we provided evidence that the near-threshold $J/\psi$ photoproduction could well be
dominated by loops with open charm hadrons. We found that the existing experimental data on $\gamma p\to J/\psi p$
can be described within the suggested mechanism through the $\Lambda_c\bar D^{(*)}$ intermediate
states if all the parameters of the model take their natural values. We identified a clear experimental
signature for this picture: The process is necessarily accompanied by the appearance of two pronounced cusps
located at the $\Lambda_c \bar D$ and the $\Lambda_c \bar D^*$ thresholds, and found the existing data
consistent with this feature within their accuracy. Since the strength of the cusps is connected to
the rate for $\gamma p\to \Lambda_c\bar D^{(*)}$, we also provided an estimate for the expected rate into
the open-charm channels and extracted the $J/\psi$-nucleon scattering length. Although all predictions
reported in this Letter should be regarded as order-of-magnitude estimates, their agreement with the
existing data on the $J/\psi$ photoproduction off the proton is remarkable. Therefore,
further experimental tests of these predictions are crucial to get a deeper understanding of the $J/\psi$
photoproduction reaction.  
The ongoing measurements of the $J/\psi$ photoproduction in Hall C at Jefferson Laboratory~\cite{Meziani:2016lhg},
which has higher statistics than GlueX, and measurements of the $\Lambda_c \bar D^{(*)}$ production will
provide crucial information.  
The prediction of the tiny $J/\psi$-nucleon scattering lengths can be tested using lattice QCD.

It should be stressed that if the open-charm loops discussed above indeed dominate the $J/\psi$-nucleon
scattering, as suggested in this Letter, the connection between the trace anomaly and the $J/\psi$-nucleon
scattering length is lost. 
This is similar to the observation that the
ratio of the decays $\psi(2S)\to J/\psi (\pi^0/\eta)$ cannot be used for an extraction of the light
quark mass ratio $m_u/m_d$ if charmed meson loops contribute to the transitions significantly~\cite{Guo:2009wr}.
This observation is intimately related to the QCD multipole expansion, that does not seem work well
in certain processes related to charmonium systems.
A further test of this physics discussed here would be a lattice calculation of the $J/\psi$-nucleon scattering
lengths, which could only be estimated in the approach used here.

\bigskip

\begin{acknowledgments}

This work is supported in part by the Chinese Academy of Sciences (CAS) under Grants No.~XDB34030303 and No.~QYZDB-SSW-SYS013, by the National Natural Science Foundation of China (NSFC) under Grants No.~11835015, No.~11947302 and No.~11961141012, by the NSFC and the Deutsche Forschungsgemeinschaft (DFG) through the funds provided to the Sino-German Collaborative Research Center ``Symmetries and the Emergence of Structure in QCD'' (NSFC Grant No. 11621131001, DFG Grant No. CRC110), and by the CAS Center for Excellence in Particle Physics (CCEPP). U.-G.M. is also supported by CAS through the President’s International Fellowship Initiative (PIFI) (Grant No.2018DM0034) and by the VolkswagenStiftung (Grant No.  93562).
Work of V.B. and A.N. was supported by the Russian Science Foundation (Grant No. 18-12-00226). Work of I.S. was supported in part by the US Department of Energy, Office of Science, Office of Nuclear Physics under Award No. DE–-SC0016583.

\end{acknowledgments}



\begin{thebibliography}{99}

\bibitem{Kharzeev:1995ij} 
  D.~Kharzeev,
  Proc.\ Int.\ Sch.\ Phys.\ Fermi {\bf 130}, 105 (1996)
  [nucl-th/9601029].



\bibitem{Kharzeev:1998bz} 
  D.~Kharzeev, H.~Satz, A.~Syamtomov and G.~Zinovjev,
  Eur.\ Phys.\ J.\ C {\bf 9}, 459 (1999)
  [hep-ph/9901375].


\bibitem{Gryniuk:2020mlh}
O.~Gryniuk, S.~Joosten, Z.~E.~Meziani and M.~Vanderhaeghen,
Phys. Rev. D \textbf{102}, 014016 (2020)
[arXiv:2005.09293 [hep-ph]].


\bibitem{Hatta:2018ina} 
  Y.~Hatta and D.~L.~Yang,
  Phys.\ Rev.\ D {\bf 98}, 074003 (2018)
  [arXiv:1808.02163 [hep-ph]].



\bibitem{Wang:2019mza} 
  R.~Wang, J.~Evslin and X.~Chen,
  Eur.\ Phys.\ J.\ C {\bf 80}, 507 (2020)
  [arXiv:1912.12040 [hep-ph]].



\bibitem{Brodsky:1989jd} 
  S.~J.~Brodsky, I.~A.~Schmidt and G.~F.~de Teramond,
  Phys.\ Rev.\ Lett.\  {\bf 64}, 1011 (1990).
  
  
\bibitem{Hosaka:2016pey}
A.~Hosaka, T.~Iijima, K.~Miyabayashi, Y.~Sakai and S.~Yasui,
PTEP \textbf{2016}, 062C01 (2016)
[arXiv:1603.09229 [hep-ph]].

\bibitem{Lebed:2016hpi}
R.~F.~Lebed, R.~E.~Mitchell and E.~S.~Swanson,
Prog. Part. Nucl. Phys. \textbf{93}, 143-194 (2017)
[arXiv:1610.04528 [hep-ph]].

\bibitem{Esposito:2016noz}
A.~Esposito, A.~Pilloni and A.~D.~Polosa,
Phys. Rept. \textbf{668}, 1-97 (2017)
[arXiv:1611.07920 [hep-ph]].

\bibitem{Guo:2017jvc}
F.-K.~Guo, C.~Hanhart, U.-G.~Mei{\ss}ner, Q.~Wang, Q.~Zhao and B.-S.~Zou,
Rev. Mod. Phys. \textbf{90}, 015004 (2018)
[arXiv:1705.00141 [hep-ph]].

\bibitem{Olsen:2017bmm}
S.~L.~Olsen, T.~Skwarnicki and D.~Zieminska,
Rev. Mod. Phys. \textbf{90}, 015003 (2018)
[arXiv:1708.04012 [hep-ph]].

\bibitem{Liu:2019zoy}
  Y.-R.~Liu, H.-X.~Chen, W.~Chen, X.~Liu and S.-L.~Zhu,
  Prog. Part. Nucl. Phys. \textbf{107}, 237-320 (2019)
  [arXiv:1903.11976 [hep-ph]].

\bibitem{Brambilla:2019esw} 
  N.~Brambilla, S.~Eidelman, C.~Hanhart, A.~Nefediev, C.-P.~Shen, C.~E.~Thomas, A.~Vairo and C.-Z.~Yuan,
  Phys.\ Rept.\  {\bf 873}, 1 (2020)
  [arXiv:1907.07583 [hep-ex]].

\bibitem{Guo:2019twa}
F.-K.~Guo, X.-H.~Liu and S.~Sakai,
  Prog. Part. Nucl. Phys. \textbf{112}, 103757 (2020)
  [arXiv:1912.07030 [hep-ph]].

\bibitem{Yang:2020atz}
  G.~Yang, J.~Ping and J.~Segovia,
  [arXiv:2009.00238 [hep-ph]].

\bibitem{Aaij:2015tga} 
  R.~Aaij {\it et al.} [LHCb Collaboration],
  Phys.\ Rev.\ Lett.\  {\bf 115}, 072001 (2015)
  [arXiv:1507.03414 [hep-ex]].



\bibitem{Aaij:2019vzc} 
  R.~Aaij {\it et al.} [LHCb Collaboration],
  Phys.\ Rev.\ Lett.\  {\bf 122}, 222001 (2019)
  [arXiv:1904.03947 [hep-ex]].



\bibitem{Ali:2019lzf} 
  A.~Ali {\it et al.} [GlueX Collaboration],
  Phys.\ Rev.\ Lett.\  {\bf 123},  072001 (2019)
  [arXiv:1905.10811 [nucl-ex]].



\bibitem{Cao:2019kst} 
  X.~Cao and J.~p.~Dai,
  Phys.\ Rev.\ D {\bf 100},  054033 (2019)
  [arXiv:1904.06015 [hep-ph]].


\bibitem{Winney:2019edt} 
  D.~Winney {\it et al.} [JPAC Collaboration],
  Phys.\ Rev.\ D {\bf 100},  034019 (2019)
  [arXiv:1907.09393 [hep-ph]].




\bibitem{Kubarovsky:2015aaa} 
  V.~Kubarovsky and M.~B.~Voloshin,
  Phys.\ Rev.\ D {\bf 92},  031502 (2015)
  [arXiv:1508.00888 [hep-ph]].



\bibitem{Karliner:2015voa} 
  M.~Karliner and J.~L.~Rosner,
  Phys.\ Lett.\ B {\bf 752}, 329 (2016)
  [arXiv:1508.01496 [hep-ph]].



\bibitem{Huang:2016tcr} 
  Y.~Huang, J.~J.~Xie, J.~He, X.~Chen and H.~F.~Zhang,
  Chin.\ Phys.\ C {\bf 40},  124104 (2016)
  [arXiv:1604.05969 [nucl-th]].



\bibitem{Meziani:2016lhg} 
  Z.~E.~Meziani {\it et al.},
  arXiv:1609.00676 [hep-ex].



\bibitem{Paryev:2018fyv} 
  E.~Y.~Paryev and Y.~T.~Kiselev,
  Nucl.\ Phys.\ A {\bf 978}, 201 (2018)
  [arXiv:1810.01715 [nucl-th]].



\bibitem{Blin:2016dlf} 
  A.~N.~Hiller Blin, C.~Fern{\'a}ndez-Ram{\'i}rez, A.~Jackura, V.~Mathieu, V.~I.~Mokeev, A.~Pilloni and A.~P.~Szczepaniak,
  Phys.\ Rev.\ D {\bf 94}, 034002 (2016)
  [arXiv:1606.08912 [hep-ph]].



\bibitem{Wang:2019krd} 
  X.~Y.~Wang, X.~R.~Chen and J.~He,
  Phys.\ Rev.\ D {\bf 99},  114007 (2019)
  [arXiv:1904.11706 [hep-ph]].



\bibitem{Wu:2019adv} 
  J.~J.~Wu, T.-S.~H.~Lee and B.~S.~Zou,
  Phys.\ Rev.\ C {\bf 100},  035206 (2019)
  [arXiv:1906.05375 [nucl-th]].



\bibitem{Cao:2019gqo} 
  X.~Cao, F.-K.~Guo, Y.~T.~Liang, J.~J.~Wu, J.~J.~Xie, Y.~P.~Xie, Z.~Yang and B.~S.~Zou,
  Phys.\ Rev.\ D {\bf 101},  074010 (2020)
  [arXiv:1912.12054 [hep-ph]].



\bibitem{Yang:2020eye} 
  Z.~Yang, X.~Cao, Y.~T.~Liang and J.~J.~Wu,
  Chin.\ Phys.\ C {\bf 44},  084102 (2020)
  [arXiv:2003.06774 [hep-ph]].



\bibitem{Paryev:2020jkp} 
  E.~Y.~Paryev,
  arXiv:2007.01172 [nucl-th].



\bibitem{Gryniuk:2016mpk} 
  O.~Gryniuk and M.~Vanderhaeghen,
  Phys.\ Rev.\ D {\bf 94}, 074001 (2016)
  [arXiv:1608.08205 [hep-ph]].



\bibitem{Boreskov:1992ur}
  K.~Boreskov, A.~Capella, A.~Kaidalov and J.~Tran Thanh Van,
  Phys. Rev. D \textbf{47}, 919 (1993).

\bibitem{Bradford:2005pt} 
  R.~Bradford {\it et al.} [CLAS Collaboration],
  Phys.\ Rev.\ C {\bf 73}, 035202 (2006)
  [nucl-ex/0509033].


  \bibitem{Bradford:2006ba}
  R.~K.~Bradford \textit{et al.} [CLAS Collaboration],
  Phys.\ Rev.\ C\ \textbf{75} (2007), 035205
  [arXiv:nucl-ex/0611034 [nucl-ex]].

  \bibitem{McCracken:2009ra}
  M.~E.~McCracken \textit{et al.} [CLAS Collaboration],
  Phys.\ Rev.\ C\ \textbf{81} (2010), 025201
  [arXiv:0912.4274 [nucl-ex]].

  \bibitem{Dey:2010hh}
  B.~Dey \textit{et al.} [CLAS Collaboration],
  Phys.\ Rev.\ C\ \textbf{82} (2010), 025202
  [arXiv:1006.0374 [nucl-ex]].

  \bibitem{Paterson:2016vmc}
  C.~A.~Paterson \textit{et al.} [CLAS Collaboration],
  Phys.\ Rev.\ C\ \textbf{93} (2016) no.6, 065201
  [arXiv:1603.06492 [nucl-ex]].


\bibitem{Seraydaryan:2013ija} 
  H.~Seraydaryan {\it et al.} [CLAS Collaboration],
  Phys.\ Rev.\ C {\bf 89}, 055206 (2014)
  [arXiv:1308.1363 [hep-ex]].

\bibitem{Dey:2014tfa}
  B.~Dey \textit{et al.} [CLAS Collaboration],
  Phys. Rev. C \textbf{89}, 055208 (2014)
  [arXiv:1403.2110 [nucl-ex]].


\bibitem{Mizutani:2017wpg} 
  K.~Mizutani {\it et al.} [LEPS Collaboration],
  Phys.\ Rev.\ C {\bf 96},  062201 (2017)
  [arXiv:1710.00169 [nucl-ex]].


\bibitem{Colangelo:2003sa} 
  P.~Colangelo, F.~De Fazio and T.~N.~Pham,
  Phys.\ Rev.\ D {\bf 69}, 054023 (2004)
  [hep-ph/0310084].



\bibitem{Guo:2010ak} 
  F.-K.~Guo, C.~Hanhart, G.~Li, U.-G.~Mei{\ss}ner and Q.~Zhao,
  Phys.\ Rev.\ D {\bf 83}, 034013 (2011)
  [arXiv:1008.3632 [hep-ph]].



\bibitem{Zyla:2020zbs} 
  P.~A.~Zyla {\it et al.} [Particle Data Group],
  PTEP {\bf 2020}, 083C01 (2020).



\bibitem{Liu:2001yx} 
  W.~Liu, C.~M.~Ko and Z.~W.~Lin,
  nucl-th/0107058.



\bibitem{Oh:2007ej} 
  Y.~Oh, W.~Liu and C.~M.~Ko,
  Phys.\ Rev.\ C {\bf 75}, 064903 (2007)
  [nucl-th/0702077].



\bibitem{Khodjamirian:2011jp} 
  A.~Khodjamirian, C.~Klein, T.~Mannel and Y.-M.~Wang,
  JHEP {\bf 1109}, 106 (2011)
  [arXiv:1108.2971 [hep-ph]].



\bibitem{Khodjamirian:2011sp} 
  A.~Khodjamirian, C.~Klein, T.~Mannel and Y.~M.~Wang,
  Eur.\ Phys.\ J.\ A {\bf 48}, 31 (2012)
  [arXiv:1111.3798 [hep-ph]].



\bibitem{Gortchakov:1995im} 
  O.~Gortchakov, M.~P.~Locher, V.~E.~Markushin and S.~von Rotz,
  Z.\ Phys.\ A {\bf 353}, 447 (1996).



\bibitem{Cheng:2004ru} 
  H.~Y.~Cheng, C.~K.~Chua and A.~Soni,
  Phys.\ Rev.\ D {\bf 71}, 014030 (2005)
  [hep-ph/0409317].



\bibitem{Strakovsky:2019bev} 
  I.~Strakovsky, D.~Epifanov and L.~Pentchev,
  Phys.\ Rev.\ C {\bf 101}, 042201 (2020)
  [arXiv:1911.12686 [hep-ph]].


\bibitem{Pentchev:2020kao} 
  L.~Pentchev and I.~I.~Strakovsky,
  arXiv:2009.04502 [hep-ph].



\bibitem{Luke:1992tm} 
  M.~E.~Luke, A.~V.~Manohar and M.~J.~Savage,
  Phys.\ Lett.\ B {\bf 288}, 355 (1992)
  [hep-ph/9204219].



\bibitem{Brodsky:1997gh} 
  S.~J.~Brodsky and G.~A.~Miller,
  Phys.\ Lett.\ B {\bf 412}, 125 (1997)
  [hep-ph/9707382].



\bibitem{Skerbis:2018lew} 
  U.~Skerbis and S.~Prelovsek,
  Phys.\ Rev.\ D {\bf 99}, 094505 (2019)
  [arXiv:1811.02285 [hep-lat]].



\bibitem{Sugiura:2017vks} 
  T.~Sugiura, Y.~Ikeda and N.~Ishii,
  EPJ Web Conf.\  {\bf 175}, 05011 (2018)
  [arXiv:1711.11219 [hep-lat]].



\bibitem{Beane:2014sda} 
  S.~R.~Beane, E.~Chang, S.~D.~Cohen, W.~Detmold, H.-W.~Lin, K.~Orginos, A.~Parre{\~n}o and M.~J.~Savage,
  Phys.\ Rev.\ D {\bf 91}, 114503 (2015)
  [arXiv:1410.7069 [hep-lat]].



\bibitem{Liu:2008rza} 
  L.~Liu, H.~W.~Lin and K.~Orginos,
  PoS LATTICE {\bf 2008}, 112 (2008)
  [arXiv:0810.5412 [hep-lat]].



\bibitem{Guo:2009wr} 
  F.-K.~Guo, C.~Hanhart and U.-G.~Mei{\ss}ner,
  Phys.\ Rev.\ Lett.\  {\bf 103}, 082003 (2009)
  [Erratum: Phys.\ Rev.\ Lett.\  {\bf 104}, 109901 (2010)]
  [arXiv:0907.0521 [hep-ph]].





\end{thebibliography}
\end{document}